\documentstyle[11pt,aaspp4]{article}

\def\spose#1{\hbox to 0pt{#1\hss}}
\def\lta{\mathrel{\spose{\lower 3pt\hbox{$\mathchar"218$}}
     \raise 2.0pt\hbox{$\mathchar"13C$}}}
\def\gta{\mathrel{\spose{\lower 3pt\hbox{$\mathchar"218$}}
     \raise 2.0pt\hbox{$\mathchar"13E$}}}

\begin{document}
\title{Galaxies or Asteroids? -- \\
12 $\mu$m Sources in 9 QSO Fields Near the Ecliptic}

\author{
A.~C.\ Quillen\altaffilmark{1}$^,$\altaffilmark{2}
}
\altaffiltext{1}{University of Arizona, Steward Observatory, Tucson, AZ 85721}
\altaffiltext{2}{E-mail: aquillen@as.arizona.edu}

\def\spose#1{\hbox to 0pt{#1\hss}}
\def\lta{\mathrel{\spose{\lower 3pt\hbox{$\mathchar"218$}}
     \raise 2.0pt\hbox{$\mathchar"13C$}}}
\def\gta{\mathrel{\spose{\lower 3pt\hbox{$\mathchar"218$}}
     \raise 2.0pt\hbox{$\mathchar"13E$}}}

\begin{abstract}
We consider 9 fields of low ecliptic latitude 
observed at 12$\mu$m centered on QSOs.
In these fields we detect 7 additional background sources
at the level of 0.4-4mJy in 70 square arcminutes. 
4 of these sources correspond to galaxies seen in the Digitized Sky Survey.  
3 of these sources are not observed in this survey.
The 4 sources with optical counterparts are low redshift objects with
low infrared to optical flux ratios and so low
inferred infrared luminosities $L_{FIR} \lta 10^9 L_\odot$.
Statistical arguments suggest that the unidentified sources are more likely
to be distant $L_{FIR} > 10^{10} L_\odot$ starbursting galaxies, rather than 
asteroids or late-type stars.

2 of our QSO fields 
are likely to have starbursting galaxies at moderately high redshift
near the QSO (in angular scale).  Since the lensing optical depth
towards these QSOs is expected to peak at moderate redshift 
there is some possibility
that these galaxies are in clusters that lense the QSOs.
If none of our sources are asteroids then the number of small (km)
sized main belt asteroids cannot be larger than 500,000 per absolute
magnitude bin.

\end{abstract}


\section {Introduction}

The ISO satellite (\cite{kes96}) 
is orders of magnitude more sensitive than IRAS at shorter
IR wavelengths ($< 20\mu$m).  
Based on the local universe IR luminosity function  (e.g. \cite{sau90},
\cite{rie86})
inferred by IRAS observations, it has been possible to make 
predictions for number counts of
objects in the mid-far IR wavelength ranges at levels fainter than
observed by IRAS.  Number counts at the 1mJy level in the mid IR 
are expected to include both normal galaxies at moderate
redshift ($z \sim 0.1$) as well as starbursts that could be 
detected out to $z\sim 1$ (e.g. \cite{fra94}, \cite{lon90}, \cite{hdf3}).

In this paper we consider background source counts in 9 QSO fields
imaged at 12$\mu$m with ISOCAM.  
The area of sky covered is larger than the ISO Hubble
Deep Field survey (\cite{hdf1}) but smaller than that being carried
out by \cite{cez97} at 25$\mu$m.
Our fields are found in the ecliptic and so can also be used to place
limits on the number of km sized asteroids in the Main Asteroid Belt.
The association with QSOs also places crude limits on magnitude 
bias for these QSOs.

\section {Observations}

Images of 9 fields centered on QSOs using the LW10 filter 
($8.6-14.4 \mu$m) were taken with ISOCAM (\cite{ces96})
(a $32\times 32$ array on the ISO satellite(\cite{kes96}) 
in the imaging mode with 6.0'' per pixel.  
Each field was observed for a 
total of 114 exposures each 5.04s long, 
taken in a $3\times 3$ raster on
the sky that offset by 18'' so that each of the nine sky positions were observed
for a total of $\sim 50$ seconds with some additional time spent on the first
position.  The total integration time per field was 9.5 minutes.
The dates and times of the observations are listed in Table 1.

Of these 114 exposures we discarded the first 30 because
the array had not yet stabilized.  After this time based
on the subsequent flat level of the sky 
we judged the array to be stabilized.
Cosmic ray glitches were removed using the 'MM'  multi-resolution
spatial and temporal routine in the CAM Interactive Analysis (CIA) package
(\footnote{CAM Interactive Analysis is a joint development by the ESA 
astrophysics division and the ISOCAM consortium 
led by the ISOCAM PI, C. Ce\'sarsky, 
Direction des Sciences de la Matie\`re, C.~E.~A.~, France}
Because of the long exposures and long total integration time, low frequency
noise was observed in each pixel as a function of time in the data
after cosmic ray removal.  To remove the low frequency noise 
we then removed third order polynomials from each pixel as a function of time.
Following this procedure, flat fields were constructed
from the images themselves 
by fitting a smooth function that varies
with time to the entire data cube.  The final
mosaic was constructed from the nine mosaic positions 
with a least-squares fit to each pixel in the images after shifting them
according to the position observed.  The final images are shown in Figure 1.

We calibrated the final images by scaling the sky value back to the original
value of the cmos ISO Automatic Analysis product which is 
estimated to be accurate within 
$\pm 30\%$ (\cite{ces96}).  Since the images used were
observed after stabilization had taken place this calibration should
be accurate to the level specified.
The resulting images have a pixel to pixel standard deviation 
of $\sim 30 \mu$Jy per $6''$ pixel in the central $25 \times 25$ pixels. 
By removing some of the low frequency noise
from the images we improved the pixel to pixel standard deviation 
by a factor of 1/5 from the cmos Automatic Analysis product.

In the center of the final images all the quasars are 
detected at the level of a few mJy. 
Because of the pointing performance of ISO
(absolute pointing error less than a few arcsec, \cite{kes96}) 
we are confident that the central sources are indeed at
the position of the QSOs.

7 additional sources (besides the QSOs) were detected in these images.
We estimate that we are complete down to the 0.3mJy level.
These sources were seen in different pixel locations
in the various frames comprising the raster 
so we are confident that these sources are not spurious
detections due to cosmic rays.  Fluxes of these 7 sources
are listed in Table 2.

Digitized Sky Survey images of the QSO field are shown in Figure 2 
with identified sources marked.  Of the 7 sources
detected 4 are galaxies resolved in the sky survey images.  
We estimate V band magnitudes and magnitude limits,  also listed in Table 2.
from the Sky Survey images by scaling from the brightness of the central
QSO.  Approximate coordinate positions of the 7 sources 
are listed in Table 3.  We judge these coordinate position to be accurate
to $\pm 6''$, the pixel size of our 12$\mu$m images.

\subsection {Number Density Estimates}
The very edge of our field has a higher level of noise than the central
region because of the effective integration
time is only 1/9-1/3 as long as in the central field.
We therefore included only the central 28 pixels in each image in our total
effective area covered.  The resulting area is 7.84 square arcminutes
per QSO field with a total of 70 square arcminutes covered in our 9 
fields.  For the four identified galaxies sources we convert this
to a number density of 100 -- 300 distant galaxies per square degree
at a limiting flux level of 0.5mJy 
(including Poisson statistical errors).  For the 3 unidentified 
sources a similar 150 objects per square degree is estimated.
Including all 7 sources we estimate a total of 300 - 500 objects per square
degree.


\section {Distant Galaxies}

In Table 2 for comparison with other IRAS samples we have computed
$L_{IR}/L_{opt}$.
We estimate $F_{opt}$ using $\nu F_\nu$ at ($0.55\mu$m or V band)
which is within $10\%$ of that estimated at estimated at 0.44$\mu$m  
for typical galaxy colors ($B-V\sim 0.5$).
We estimate $F_{IR}$ as
$\nu F_\nu$ at $60\mu$m using a conversion
of $F_{\nu(60\mu{rm m}} / F_{\nu(12\mu{\rm m}} = 16$ based on the mean
value of the IRAS bright galaxy sample (\cite{soi89}).  
This value of $L_{IR}$ is approximately equal to the parameter
$L_{FIR}$ of \cite{hel88} based on the 60 and 100$\mu$m flux for
a galaxy with a mean $F_{60\mu{\rm m}}/F_{100\mu{\rm m}} = 0.5$.
For field galaxies no correlation between $L_{IR}$ and 
$F_{60\mu {\rm m}}/F_{25\mu{\rm m}}$ was observed (\cite{rie86}) 
however there could 
be correlation between $F_{60\mu{\rm m}}/F_{12\mu{\rm m}}$ and
$L_{IR}$ because of the differing contributions of cirrus and star forming
regions to the IR emission (\cite{hel86}) in quiescent and starbursting 
galaxies.

From Table 2 we can see that for the 4 galaxies
with bright optical counterparts, log$(L_{IR}/L_{opt}) < 0.0$.
Based on the correlation between $L_{IR}/L_{opt}$ with
$L_{IR}$ (\cite{soi89})
this suggests that these galaxies are low luminosity 
$L_{IR} \sim 10^9 L_\odot$ galaxies (with $z<0.1$).
On the other hand 
the limiting flux ratios of the 3 12$\mu$m sources without optical 
counter parts are high enough that if they are galaxies 
they would have high IR luminosities $L_{IR} > 10^{10} L_\odot$
and therefore would be more distant ($z\sim 1$). 
If these sources are faint galaxies they should 
be visible in optical images deeper than the digitized sky survey 
images.

If all 7 of our 12$\mu$m sources are galaxies then our total
number density is somewhat higher (a factor of 2-4 higher)
than predicted by \cite{fra91}
though within small number statistics our number density is 
similar to that predicted by \cite{rie97}.  
However the optical to IR color ratios
differ from those predicted by these models 
for the 4 sources with optical counterparts.
Using the local universe luminosity function of \cite{sau90}
and the distribution of $F_{12\mu {\rm m}} /F_{60 \mu {\rm m}}$
from \cite{soi89} we find that for sources at a limiting flux
of 0.4mJy the mean luminosity is $L_{IR} \sim 10^{10}L_\odot$ and
is larger than
we infer from the optical to 12$\mu$m colors for the 4 sources
with optical counter parts.  However this discrepancy could be
alleviated if there is a dependence of 
$F_{12\mu {\rm m}} /F_{60 \mu {\rm m}}$ on $L_{IR}$, which is not 
unexpected if the emission from low luminosity galaxies is dominated by cirrus
emission (\cite{hel86}).

If the 3 objects without counterparts are starbursting galaxies
then as shown by \cite{hdf3}, \cite{hdf4} and \cite{hdf5} 
cosmological models with evolution are required to explain
the number counts.  This is not unexpected based on the large number
of high redshift starbursting galaxies observed in the Hubble Deep Field.

\subsection {Magnitude or lensing bias and the possibility of 
association with with QSO}

It is unlikely that any of our sources are at the QSO redshift
since they are close in flux level to the QSO itself and such bright 
implied galaxies luminosities in the QSO environment are unlikely.
The QSO's redshifts are $z=1.5-2$ so that the lensing
optical depth towards them peaks at about $z=1$.  This makes it unlikely
that a lensing cluster at low redshift is responsible for
both brightening the QSO and containing a galaxy 
detectable at 12$\mu$m.  
Spiral galaxies, being much more numerous and brighter
in the mid-IR than ellipticals are more likely to be detected at 12$\mu$m
and are unlikely to be in clusters.
However it is possible that the 3 sources without optical counterparts
are starbursting galaxies  and are  part of a 
cluster which could be lensing the QSOs.

\section {Other Possibilities for the 3 Sources Without Optical Counterparts}

\subsection {The possibility of Galactic sources or M stars}
Our fields are all at high galactic latitudes,
greater than $55 ^\circ$ from the galactic plane 
except for POX 42 which
is at a galactic latitude of $42^\circ$.
At these high galactic latitudes,
most galactic sources are too scarce to be observed at the 
number density that we observe them at the 0.5 mJy level
at 12$\mu$m.

%

Here we consider the possibility that we could be detecting late M stars
which are quite numerous.
Using absolute V magnitudes from \cite{leg92} and V-[12] colors from
\cite{ken95}  we estimate that an M0, M4 star would be detected in 12$\mu$m at 
the 0.5 mJy level to a distance of only 300 and 50 pc respectively.
Summing the entire luminosity function (M0-M8) of \cite{gou96} over the volume 
limit given by the 0.5 mJy 12$\mu$m detection limit gives a total
of $\sim 20$ sources expected per degree which is insufficient in number
for us to expect the 3 sources we see in 70 square arcminutes.
We therefore find that it is very unlikely that our 3 unidentified sources
are late-type stars.  We have not used the optical to 12$\mu$m color limits
placed from the digitized sky survey because  nearby stars
could have moved since the sky survey images were observed.

According to the models of \cite{bur97} a 1 Gyr old brown dwarf
with a mass of 40 Jupiter masses would only be seen at a level of 0.5 mJy
at a distance of 14 pc.  A rather high 
total number density of 700 bright brown dwarfs per pc$^3$
would be required so that we could detect them in our images.  It's
very unlikely that our unidentified objects are brown dwarfs.

\subsection {The possibility of asteroids}

All of our fields are at low ecliptic latitude;
within $10^\circ$ of the ecliptic except
POX 42 (at $18^\circ$) and 0059-2735 (at $27^\circ$) from the ecliptic.
Here we consider the likelihood of detecting Main Belt asteroids.  

None of our unidentified
objects are numbered asteroids listed currently
in the database provided by Minor Planet Center  
(operated at the Smithsonian Astrophysical Observatory, 
under the auspices of Commission 20 of the International Astronomical Union).
Asteroids with known orbit elements, numbering about 35000,
(The Asteroid Orbital Elements Database of Lowell Observatory)
are only complete to about $H=13$ absolute magnitude (where the absolute
magnitude is computed at 1AU from both the observe and the sun 
in V band, \cite{kui58}), and would be much brighter than 1mJy at 12$\mu$m
if they were in the Main Belt (less than 4 AU from the sun).  

We therefore must consider deeper asteroid surveys.
As part of the Spacewatch Survey (\cite{geh91}),
\cite{jed97} have completed a deep ($V < 21$) survey of 
the Main Asteroid Belt \cite{jed97} which is complete to $H \sim 17$.
These authors find approximately 50000 asteroids per absolute magnitude
bin in each of 3 regions.  Their 
inner, mid and outer regions are centered at 2.25,2.75 and 3.25 AU
respectively from the sun and are 0.5 AU wide.
Using the standard thermal model, and a bond emissivity
of 0.1, we estimate that asteroids of $H=$19,18, and 17 
(having sizes of $\sim$ 0.8,1.3, and 2.0 km) would be detected at opposition
at the 1mJy level in 12$\mu$m in the inner, mid and outer regions respectively.
\cite{jed97} found that the numbers of asteroids does not depend
strongly on the absolute magnitude for $15<H<17$.
Using the number of asteroids measured at H=17 mag in their last
complete magnitude bin and summing over the 3 regions 
we compute a total of 20 objects per square degree would
be detected within $10^\circ$
of the ecliptic at 12$\mu$m at the 1mJy level from the Main Belt (distributing
the asteroids evenly within $10^\circ$ of the ecliptic).
It is therefore unlikely that all 3 of our sources are asteroids.

\section {Discussion}
We find that it most likely that our 7 additional sources are distant 
galaxies.
The estimated number densisty of objects is somewhat higher than
predicted from the local IRAS derived luminosity functions,  however
with additional star formation at high redshift our 
three sources with no bright optical counterparts
are likely to be distant starbursting galaxies.  The 4 sources
with couterparts are only predicted with the local IRAS luminosity function
if the 12$\mu$m  flux is relatively high in low $L_{IR}$ galaxies
compared to starbursting galaxies, which is not 
unexpected since these galaxies should be dominated by emission from
cirrus at 12 $\mu$m.

2 of our QSO fields (the ones that have 12$\mu$m sources 
with no bright optical counterparts) 
are likely to have starbursting galaxies at moderately high redshift
near the QSO (in angular scale).  Since the lensing optical depth
towards these QSOs is expected to peak at $z\sim 1$ there is some possibility
that these galaxies are in clusters that lense the QSOs.
If none of our sources are asteroids then the number of small (km) 
sized main belt asteroids cannot be larger than 500,000 per absolute
magnitude bin.

\vskip 0.5truein
\acknowledgments

The Infrared Space Observatory (ISO) is an ESA project funded
by ESA Member States (especially the PI countries: France, Germany, 
The Netherlands, and the United Kingdom
with the participation of ISAS and NASA).
We thank D. van Buren, M. Seh, K. Ganga, R. Hurt, 
L. Hermans and the ISO team at 
IPAC for help with the data reduction of the ISOCAM images.

I thank M. Sykes for his code that predicts 12$\mu$m fluxes
of asteroids, and teaching me alot about about asteroids.  
We acknowledge helpful discussions and correspondence with 
R. Jedicke, M. Rieke, G. Rieke, P. Hall, D. Hines, R. Kennicutt, P. Massey, 
I. Hook, A. Gould, M. Sykes, C. Neese, T. Fleming, M. Hanson, R. Hurt, and
K. Luhman.
We also acknowledge support from NSF grant AST-9529190 
and NASA project no. NAG-53359.

Figure 2 and Table 2 made use of the Digitized Sky Survey.
The Digitized Sky Surveys were produced at the Space Telescope 
Science Institute
under U.S. Government grant NAG W-2166. The images of these surveys are
based on photographic data obtained using the Oschin Schmidt Telescope on
Palomar Mountain and the UK Schmidt Telescope. The plates were processed 
into the present compressed digital form with the permission of these 
institutions. 
 The National Geographic Society - Palomar Observatory Sky Atlas (POSS-I) 
was made by the California Institute of Technology with grants 
from the National Geographic Society. 
  The Second Palomar Observatory Sky Survey (POSS-II) was made by the
California Institute of Technology with funds from the National Science 
Foundation, the National Geographic Society, the Sloan Foundation, the 
Samuel Oschin Foundation, and the Eastman Kodak Corporation. 
     The Oschin Schmidt Telescope is operated by the California Institute 
of Technology and Palomar Observatory. 
     The UK Schmidt Telescope was operated by the Royal Observatory Edinburgh,
with funding from the UK Science and Engineering Research Council 
(later the UK Particle Physics and Astronomy 
Research Council), until 1988 June, and thereafter
by the Anglo-Australian Observatory. The blue plates of the southern Sky Atlas
and its Equatorial Extension (together known as the SERC-J), as well as the
Equatorial Red (ER), and the Second Epoch [red] Survey (SES) were all
taken with the UK Schmidt. 

The research and computing needed to generate 
The Asteroid Orbital Elements Database 
were funded principally by NASA grant NAGW-1470, and in part by the Lowell
Observatory endowment.   
This database is created and made available by T. Bowell.

{}

%

\begin{deluxetable}{lccccccccccc}
\footnotesize
\tablecaption{Dates and Times of Observations}
\tablewidth 4.0truein
\tablehead{
\colhead{QSO Field}      &
\colhead{Date (UT)}      &
\colhead{Start Time}     & 
\colhead{End  Time }      
}
\startdata
POX 42    & 28/07/96 & 05:43:34 & 05:53:31 \nl
UM 275    & 12/12/96 & 21:48:52 & 21:59:11 \nl
UM 208    & 06/12/96 & 19:51:25 & 20:01:49 \nl
0059-2735 & 30/11/96 & 03:59:32 & 04:09:53 \nl
1246-0542 & 28/07/96 & 06:52:36 & 07:03:01 \nl
UM 288    & 12/12/96 & 18:31:34 & 18:41:48 \nl
1309-0536 & 30/07/96 & 18:28:34 & 18:38:38 \nl
1331-0108 & 03/08/96 & 02:10:58 & 02:21:04 \nl
\enddata
\tablenotetext{}{
NOTES.--
Dates are given as day/month/year.
}
\end{deluxetable}

\begin{deluxetable}{llcccccccccc}
\footnotesize
\tablecaption{Source Fluxes and Offsets}
\tablehead{
\colhead{Source$^a$}       &
\colhead{QSO Field}            &
\colhead{Offset $x ('')^b$}          & 
\colhead{Offset $y ('')^b$}          & 
\colhead{$F_{12\mu{\rm m}}$}          &
\colhead{$V$ (mag)$^c$}              &
\colhead{log $F_{IR}/F_{opt}^e$}        
} 
\startdata
%
1 & POX 42    &  87.6 &   21.5 &  $1.01 \pm 0.12$ & 18.1    & $-0.18$ \nl
2 & UM 275    &  29.1 &  -84.7 &  $3.77 \pm 0.18$ & $>$20.3 & $>1.3$ \nl
3 & UM 275    & -84.0 &  -10.1 &  $0.48 \pm 0.11$ & 18.1    & $-0.53$  \nl
4 & UM 275    &  37.3 &   55.2 &  $0.41 \pm 0.10$ & $>$20.3 & $>0.3$ \nl
5 & UM 208    &  32.4 &  -23.1 &  $0.61 \pm 0.10$ & 16.1    & $-1.16$  \nl
6 & UM 208    & -20.0 &  -84.6 &  $0.49 \pm 0.10$ & $>$20.5 & $>0.5$ \nl
7 & 0059-2735 &  43.9 &  -55.3 &  $0.77 \pm 0.10$ & 16.8    & $-0.82$  \nl
\enddata
\tablenotetext{}{
NOTES.--  
$^a$ We assign numbers to the detected sources.
$^b$ Positions of source from the QSO where a positive $x$,$y$ offset is to the 
west, north of the QSO respectively. 
Positions are accurate to $\pm 6''$.
$^c$ V mag estimated from the Digitized Sky Survey images (see Figure 2) 
by scaling from the QSO V band magnitude.  
Values preceded with $>$ are detection limits
for sources that were not detected in the Digitized Sky Survey.
$^d$ Flux ratio or limit from the Digitized sky survey?
$^e$ See text. 
}
\end{deluxetable}

\begin{deluxetable}{lccccccccccc}
\footnotesize
\tablecaption{Astrometry of Sources}
\tablewidth 3.0truein
\tablehead{
\colhead{Source}       &
\colhead{RA(1950)}       &
\colhead{DEC(1950)}      
}
\startdata
1 & 11:58:05.8 & $-$18:42:42 \nl
2 & 00:43:37.6 & $+$00:46:38 \nl
3 & 00:43:45.1 & $+$00:47:53 \nl
4 & 00:43:37.0 & $+$00:48:58 \nl
5 & 00:07:40.6 & $-$00:04:39 \nl
6 & 00:07:44.1 & $-$00:05:40 \nl
7 & 00:59:49.8 & $-$27:36:52 \nl
\enddata
\tablenotetext{}{
NOTES.--    
Positions are accurate to $\pm 6''$.
}
\end{deluxetable}

\end{document}